\documentclass{PoS}

\title{Multi-epoch VLBI of a double maser super burst}

\ShortTitle{Double Maser Super Burst}

  \author{\speaker{Ross A. Burns}$^{ab}$, Olga Bayandina$^{c}$, Gabor Orosz$^{de}$, Mateusz Olech$^{f}$, Katharina Immer$^{a}$, Jay Blanchard$^{a}$, Benito Marcote$^{a}$, Huib van Langevelde$^{a}$, Tomoya Hirota$^{b}$, Kee-Tae Kim$^{g}$, Irina Val`tts$^{c}$, Nadya Shakhvorostova$^{c}$, Georgij Rudnitskii$^{h}$, Alexandr Volvach$^{ij}$, Larisa Volvach$^{ij}$, Gordon MacLeod$^{k}$, James O. Chibueze$^{k}$, Gabriele Surcis$^{l}$, Busaba Kramer$^{m}$, Willem Baan$^{en}$, Crystal Brogan$^{o}$, Todd Hunter$^{o}$ and Stan Kurtz$^{p}$\\
     \llap{$^a$}Joint Institute for VLBI ERIC,
     Oude Hoogeveensedijk 4, 7991 PD Dwingeloo, The Netherlands\\
     \llap{$^b$}Mizusawa VLBI Observatory, National Astronomical Observatory of Japan,
      2-21-1 Osawa, Mitaka, Tokyo 181-8588, Japan\\
     \llap{$^c$}Astro Space Center, Lebedev Physical Institute,
     Russian Academy of Sciences, Leninskiy Prospekt 53, Moscow 119333, Russia\\   \llap{$^d$}School of Natural Sciences, University of Tasmania, Private Bag 37, Hobart, Tasmania 7001, Australia\\
     \llap{$^e$}Xinjiang Astronomical Observatory, Chinese Academy of Sciences, 150 Science 1-Street, Urumqi, Xinjiang 830011, China\\
     \llap{$^f$}Centre for Astronomy, Faculty of Physics, Astronomy and Informatics, Nicolaus Copernicus University, Grudziadzka 5, 87-100 Torun, Poland\\
     \llap{$^g$}Korea Astronomy and Space Science Institute 776, Daedeokdae-ro, Yuseong-gu, Daejeon, 34055, Republic of Korea\\
     \llap{$^h$}Lomonosov Moscow State University, Sternberg Astronomical Institute, Moscow, 119234 Russia\\
     \llap{$^i$}Radio Astronomy and Geodinamics Department of Crimean Astrophysical Observatory, Katsively, RT-22 Crimea\\
     \llap{$^j$}Institute of Applied Astronomy, Russian Academy of Sciences, St. Petersburg, 191187 Russia\\
     \llap{$^k$}Hartebeesthoek Radio Astronomy Observatory, PO Box 443, Krugersdorp 1740, South Africa\\
     \llap{$^l$}INAF Osservatorio Astronomico di Cagliari, Via della Scienza 5, 09047 Selargius, Italy\\
     \llap{$^m$}Max-Planck-Institut f{\"u}r Radioastronomie, Auf dem H{\"u}gel 69,
53121 Bonn, Germany\\
     \llap{$^n$}Netherlands Institute for Radio Astronomy, Dwingeloo, The Netherlands\\
     \llap{$^o$}NRAO, 520 Edgemont Road, Charlottesville, VA 22903, USA\\
     \llap{$^p$}Instituto de Radioastronom{\'\i}a y Astrof{\'\i}sica, Universidad Nacional Aut$\acute{o}$noma de M$\acute{e}$xico, Apartado Postal 3-72, Morelia 58089, M$\acute{e}$xico\\
     E-mail:  \email{ross.burns@nao.ac.jp}}

\abstract{In a rare and spectacular display, two well-known massive star forming regions, W49N and G25.65+1.05, recently underwent maser 'super burst' - their fluxes suddenly increasing above 30,000 and 18,000 Jy, respectively, reaching several orders of magnitude above their usual values. 
In quick-response, ToO observations with the EVN, VLBA and KaVA were obtained constituting a 4 week campaign - producing a high-cadence multi-epoch VLBI investigation of the maser emission. The combination of high-resolution, polarisation and flux monitoring during the burst provides one of the best accounts, to date, of the maser super burst phenomenon, aiding their use as astrophysical tools. These proceedings contain the preliminary results of our campaign.}

\FullConference{14th European VLBI Network Symposium \& Users Meeting (EVN 2018)\\
		8-11 October 2018\\
		Granada, Spain}

\begin{document}

\begin{figure}[h!]
\begin{center}
\hspace{-1cm}
\includegraphics[scale=0.4]{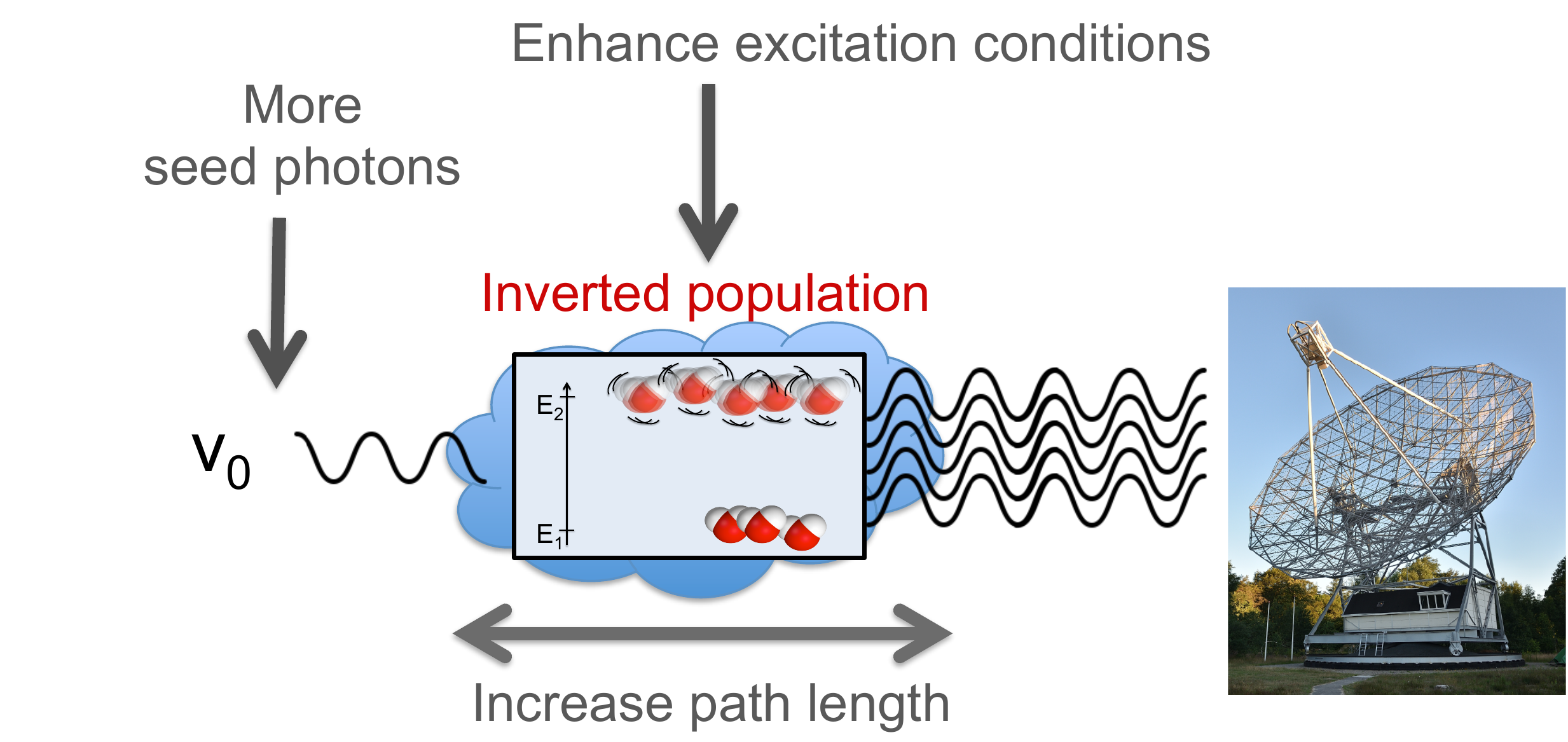}
\caption{Illustration of the main factors influencing the observed maser flux. Photo credit: Paul Boven
\label{CrappyCrap}}
\end{center}
\end{figure}

\section{Introduction}
The maser `super burst' describes an event where a maser emitting region exhibits a sudden increase in flux density of several orders of magnitude in the space of months, weeks, days, or even hours (\emph{ex.} \cite{Matveenko88,Garay89,Shimoikura05,Hirota14b,Volvach19}). Currently there are 5 recognised super burst star forming regions (SFRs): Orion KL, NGC 6334I, S255, W49N and G25.65+1.05.

Super bursts constitute rare, transient events and consequently their mechanism of action remains enigmatic. 
The widely cited Deguchi \& Watson model \cite{DnW89} generally allows for three avenues of flux enhancement: an increase in the incident continuum photon flux entering the maser; mechanical/radiative induction of more favourable maser pumping conditions; or an increase in the path length of the maser along the line of sight to the observer. These three scenarios are illustrated in the schematic diagram in Figure~\ref{CrappyCrap}. Using observations of the super burst phenomenon we hope to distinguish which of these processes is taking place in the observed maser sources.

During the IAU symposium 336 in September 2017 two well-known massive star forming regions, W49N and G25.65+1.05, simultaneously underwent maser super bursts - their fluxes increasing several orders of magnitude above their usual values. These proceedings present early results from an observing campaign investigating the mechanism of these maser super bursts.

\section{Observations}
 Burst activity of $>18,000$ Jy was reported by the single-dish Maser Monitoring Organisation (M2O) on the 7-8th of September, after which the maser rose to 65,000 Jy in late September \cite{Volvach17ATEL,Volvach19}.
Target of opportunity observations of the 22 GHz water maser in G25.65+1.05 were requested to the European VLBI network (EVN), the KVN and VERA Array (KaVA) and the Very Long Baseline Array (VLBA) and observations were conducted on the 2nd, 11th and 28th of October 2017, respectively. Data were correlated by the representative correlators for each array and processing was conducted with AIPS (Astronomical Image Processing System) details of which will be presented in a forthcoming series of papers. Maser self-calibration was performed without phase referencing to a continuum source, thus the maps center on a common reference maser.

\begin{figure}[h!]
\begin{center}
\hspace{-0.44cm}
\includegraphics[scale=1.21]{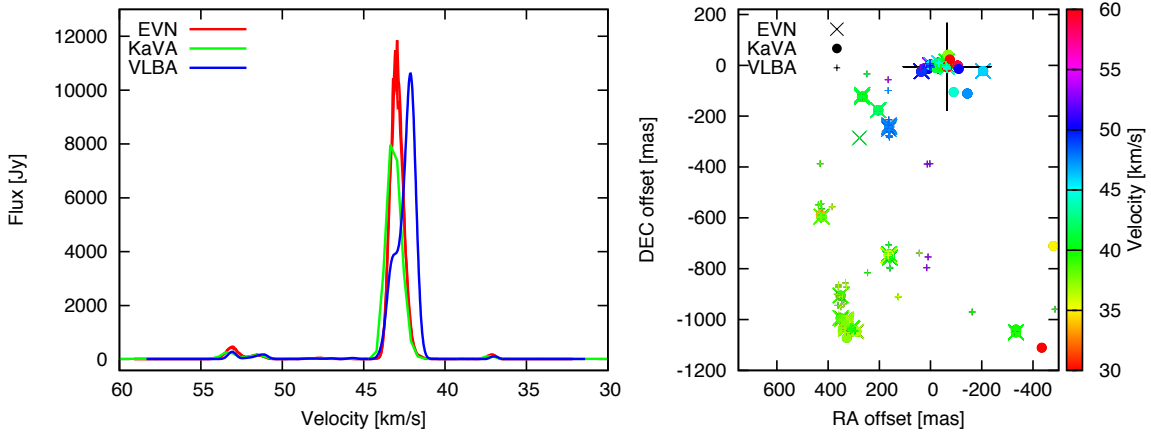}
\caption{\emph{Left} shows the scalar averaged spectrum of maser emission in G25.65+1.05 for each of the arrays for observing dates: 2nd October (EVN), 11th October (KaVA), 28th October (VLBA). \emph{Right} shows the distribution of maser spots in G25.65+1.05 where colour scale indicates LSR velocity. A black cross indicates the position of the bursting maser feature.
\label{ALL}}
\end{center}
\end{figure}

\section{Discussion}

A preliminary maser spotmap is shown in Figure~\ref{ALL}.
Maser emission is seen to form two arc structures at \emph{(RA,DEC)=(0,0)} and \emph{(RA,DEC)=(350,-1000) mas offset}. These are both sites of continuum emission revealed by JVLA observations by Bayandina et al. (in prep) and are thought to be regions of ionised gas either of a stellar origin or relating to feedback-driven shocks. Such shocks likely provide the collisional pumping energy required to induce maser action. The super burst maser resides in the center of an arc of masers that extends E-W, resembling a shock surface.

The flux density of the bursting feature is seen to vary by several hundreds of Jy during the observation campaign - indeed G25.65+1.05 is known to be extremely variable \cite{Volvach19}. It can also be noted from the spectrum in Figure~\ref{ALL} that significant flux variation is seen exclusively in the bursting maser spectral feature; the low-flux spectral features remain constant during the campaign. Thus, the exclusivity of the flux variations to one spectral (and spatial) component requires whatever mechanisms to be driving the super burst to be strictly localised to a sub-milliarcsecond region.

A comparison of the simultaneous total flux measured by single dish instruments, with the VLBI cross correlation flux indicates the compactness of the maser emission. During the VLBI campaign, VLBI flux recovery of up to 100\% was observed (during the VLBA measurement) indicating highly compact emission. Furthermore, the masers in G25 were shown to be unsaturated \cite{Volvach19}, which is a fundamental requirement for non-linear flux enhancement in the maser overlap scenario.

Taken together, the observational data strongly favor the explanation of the super burst in G25.65+1.05 to have been caused by the overlapping of maser features along the line of sight to the observer. Further analysis of the observational data including structural analyses, proper motions and observations with various other instruments will be able to further clarify the origin of the 22 GHz water maser super burst in G25.65+1.05.

\subsection{W49N and outlook for the M2O-VLBI}
Due to the rapid cessation of the super burst in W49N our observations were too late to capture the phenomenon as it happened. The data taken will be used to establish a current observational basis for future burst events. 

Our research group, the M2O-VLBI, have taken steps to ensure that future maser bursts identified by single-dish observations may be promptly observed with VLBI via a triggerable observation campaign.



\end{document}